\documentstyle[epsfig]{prhep97}

\makeatletter
\let\chapter\hid@chapter
\makeatother

\def\beq{\begin{equation}}
\def\eeq{\end{equation}}
\def\beqar{\begin{eqnarray}}
\def\eeqar{\end{eqnarray}}
\def\barr#1{\begin{array}{#1}}
\def\earr{\end{array}}
\def\bfi{\begin{figure}}
\def\efi{\end{figure}}
\def\btab{\begin{table}}
\def\etab{\end{table}}
\def\bce{\begin{center}}
\def\ece{\end{center}}

\def\text{\textstyle}

\def\de{\delta}

\def\De{\Delta}

\def\reffi#1{\mbox{Fig.~\ref{#1}}}

\def\citere#1{\mbox{Ref.~\cite{#1}}}

\def\mathswitchr#1{\relax\ifmmode{\mathrm{#1}}\else$\mathrm{#1}$\fi}

\newcommand{\PW}{\mathswitchr W}
\newcommand{\PZ}{\mathswitchr Z}

\newcommand{\Pb}{\mathswitchr b}

\newcommand{\Pt}{\mathswitchr t}

\def\mathswitch#1{\relax\ifmmode#1\else$#1$\fi}

\newcommand{\MW}{\mathswitch {M_\PW}}

\newcommand{\MZ}{\mathswitch {M_\PZ}}

\newcommand{\Mb}{\mathswitch {m_\Pb}}
\newcommand{\Mt}{\mathswitch {m_\Pt}}

\newcommand{\scrs}{\scriptscriptstyle}
\newcommand{\sw}{\mathswitch {s_{\scrs\PW}}}

\newcommand{\cw}{\mathswitch {c_{\scrs\PW}}}

\newcommand{\GF}{\mathswitch {G_\mu}}

\def\tb{\tan\beta}

\newcommand{\ses}{self-en\-er\-gies}

\renewcommand{\Re}{\mathop{\mathrm{Re}}}

\def\draftdate{\relax}
\def\mda{\relax}
\def\mua{\relax}
\def\mla{\relax}
\def\draft{
\def\thtystars{******************************}
\def\sixtystars{\thtystars\thtystars}
\typeout{}
\typeout{\sixtystars**}
\typeout{* Draft mode!
         For final version remove \protect\draft\space in source file
*}
\typeout{\sixtystars**}
\typeout{}
\def\draftdate{\today}
\def\mua{\marginpar[\boldmath\hfil$\uparrow$]%
                   {\boldmath$\uparrow$\hfil}%
                    \typeout{marginpar: $\uparrow$}\ignorespaces}
\def\mda{\marginpar[\boldmath\hfil$\downarrow$]%
                   {\boldmath$\downarrow$\hfil}%
                    \typeout{marginpar: $\downarrow$}\ignorespaces}
\def\mla{\marginpar[\boldmath\hfil$\rightarrow$]%
                   {\boldmath$\leftarrow $\hfil}%
                    \typeout{marginpar:
$\leftrightarrow$}\ignorespaces}
\def\Mua{\marginpar[\boldmath\hfil$\Uparrow$]%
                   {\boldmath$\Uparrow$\hfil}%
                    \typeout{marginpar: $\Uparrow$}\ignorespaces}
\def\Mda{\marginpar[\boldmath\hfil$\Downarrow$]%
                   {\boldmath$\Downarrow$\hfil}%
                    \typeout{marginpar: $\Downarrow$}\ignorespaces}
\def\Mla{\marginpar[\boldmath\hfil$\Rightarrow$]%
                   {\boldmath$\Leftarrow $\hfil}%
                    \typeout{marginpar:
$\Leftrightarrow$}\ignorespaces}
\overfullrule 5pt
\oddsidemargin -15mm
\marginparwidth 29mm
}

\begin{document}

\thispagestyle{empty}

\null
\hfill KA-TP-25-1997\\
\null
\hfill hep-ph/9712226\\
\vskip .8cm
\begin{center}
{\Large \bf QCD Corrections to Electroweak Precision Observables\\[.5em]
in SUSY Theories%
\footnote{Contribution to the proceedings of the {\em International 
Europhysics Conference on High-Energy Physics}, Jerusalem, Israel, 
August 19--26, 1997.}
}
\vskip 2.5em
{\large
{\sc Georg Weiglein}\\[1ex]
{\normalsize \it Institut f\"ur Theoretische Physik, Universit\"at
Karlsruhe,
D-76128 Karlsruhe, Germany}
}
\vskip 2em
\end{center} \par
\vskip 1.2cm
\vfil
{\bf Abstract} \par
The two-loop QCD corrections to the $\rho$~parameter are derived in the
Minimal Supersymmetric Standard Model. They turn out to be sizable and
modify the one-loop result by up to 30\%. Furthermore exact results for
the gluonic corrections to $\Delta r$ are presented and compared with
the leading contribution entering via the $\rho$~parameter.
\par
\vskip 1cm
\null
\setcounter{page}{0}
\clearpage

\authorrunning{G.\,Weiglein}
\titlerunning{{\talknumber}: QCD Corrections to Precision Observables
in SUSY}
 

\def\talknumber{1106} 

\title{{\talknumber}: QCD Corrections to Electroweak Precision
Observables in SUSY Theories}
\author{Georg\,Weiglein
(georg@itpaxp5.physik.uni-karlsruhe.de)}
\institute{Institut f\"ur Theoretische Physik, Universit\"at Karlsruhe,
76128 Karlsruhe, Germany}

\maketitle

\begin{abstract}
The two-loop QCD corrections to the $\rho$~parameter are derived in the
Minimal Supersymmetric Standard Model. They turn out to be sizable and
modify the one-loop result by up to 30\%. Furthermore exact results for
the gluonic corrections to $\Delta r$ are presented and compared with
the leading contribution entering via the $\rho$~parameter.
\end{abstract}
\section{Introduction}

The Minimal Supersymmetric Standard Model (MSSM) provides the most
predictive framework beyond the Standard Model (SM). While the
direct search for supersymmetric particles has not been successful yet,
the precision tests of the theory provide the possibility for
constraining the parameter space of the model and could eventually
allow to distinguish between the SM and Supersymmetry via their
respective
virtual effects. 
While the SM predictions for $\De r$ and the Z-boson observables include
leading terms at two-loop and three-loop order, the corresponding
predictions within
the MSSM have been restricted so far to one-loop order~\cite{susyprec}.
In order to treat the MSSM at the same level of
accuracy as the SM, higher-order contributions should be incorporated.
In this paper results for the QCD corrections to the $\rho$~parameter
in the MSSM are presented~\cite{susydelrhos,susydelrhol}. In addition,
the result
for the gluonic contribution to $\De r$ is derived and compared
with the approximation based on the contribution entering via the
$\rho$~parameter.

\section{QCD corrections to the $\rho$~parameter in the MSSM}

In the MSSM,
the leading contributions of scalar quarks to $\De r$ and the leptonic
Z-boson observables enter via the $\rho$~parameter.
The contribution of squark
loops to the $\rho$~parameter can be written in terms of the transverse
parts of the W- and Z-boson \ses\ at zero momentum-transfer,
\beq
\Delta \rho =
\frac{\Sigma^{\PZ}(0)}{\MZ^2} - \frac{\Sigma^{\PW}(0)}{\MW^2} .
\eeq
The one-loop result for the stop/sbottom doublet in the MSSM
reads~\cite{R6rho}
$$
\Delta \rho ^{\rm SUSY}_0 = \frac{3 \GF}{8 \sqrt{2} \pi^2} 
\left[ - sc
F_0( m_{\tilde{\Pt}_1}^2,  m_{\tilde{\Pt}_2}^2 )
+ c
F_0( m_{\tilde{\Pt}_1}^2, m_{\tilde{\Pb}_L}^2 ) 
+ s
F_0(m_{\tilde{\Pt}_2}^2,  m_{\tilde{\Pb}_L}^2 ) \right] ,
$$
where $s = \sin^2 \theta_{\tilde{\Pt}}$, $c = \cos^2
\theta_{\tilde{\Pt}}$, 
$\theta_{\tilde{\Pt}}$ is the stop mixing angle, 
and mixing in the sbottom sector has been neglected.
The function $F_0(x,y)$ has the form
$
F_0(x,y)= x+y - \frac{2xy} {x-y} \log \frac{x}{y}  .
$
It vanishes if the squarks are degenerate in 
mass.
In the limit of a large mass splitting between the
squarks it is proportional to the heavy squark mass 
squared.
This is in analogy to the case of the top/bottom
doublet in the SM~\cite{velt},
$
\Delta \rho_0^{\rm SM} = \frac{3 \GF}{8 \sqrt{2} \pi^2}
F_0(\Mt^2,\Mb^2) \approx
\frac{3 \GF \Mt^2}{8 \sqrt{2} \pi^2} .
$

Since the contribution of a squark doublet vanishes
if all masses are degenerate, in most SUSY scenarios only the third
generation contributes. 
In the third generation the top-quark mass enters the mass matrix of
the scalar partners of the top quark and can give rise to a large
mixing in the stop sector and to a large splitting between the stop and
sbottom masses.

The two-loop Feynman diagrams of the
squark loop contributions to $\De\rho$ at ${\cal O}(\alpha \alpha_s)$
can be divided into diagrams in which a gluon is exchanged, into
diagrams with gluino exchange, and into pure scalar diagrams.
After the inclusion of the corresponding counterterms the
contribution of the pure scalar diagrams vanishes and the other two
sets are separately ultraviolet finite and gauge-invariant
(see \citere{susydelrhol}).

The result for the gluon-exchange contribution is given by a simple
expression resembling the one-loop result 
$$
\Delta \rho ^{\rm SUSY}_{1, {\rm gluon}} =
\frac{\GF \alpha_s}{4 \sqrt{2} \pi^3} \left[- sc 
F_1( m_{\tilde{\Pt}_1}^2,  m_{\tilde{\Pt}_2}^2) 
+ c F_1 ( m_{\tilde{\Pt}_1}^2, m_{\tilde{\Pb}_L}^2 )
+ s F_1 (m_{\tilde{\Pt}_2}^2,  m_{\tilde{\Pb}_L}^2 ) \right] .
$$
The two-loop function $F_1(x,y)$ is given in terms of
dilogarithms by
\beqar
F_{1}(x,y) &=& x+y- 2\frac{xy}{x-y} \log \frac{x}{y} \left[2+
\frac{x}{y} \log \frac{x}{y} \right] \nonumber \\
&& {} +\frac{(x+y)x^2}{(x-y)^2}\log^2 \frac{x}{y}
-2(x-y) {\rm Li}_2 \left(1-\frac{x}{y} \right) .
\eeqar
It is symmetric in the interchange of $x$ and $y$ and
vanishes for degenerate masses, $F_1(m^2,m^2)=0$, while in the case
of
large mass splitting it increases with the heavy scalar quark mass
squared: $F_1 (m^2,0) = m^2( 1 +\pi^2/3)$.

The gluon-exchange contribution is of the order of 10--15\% of the
one-loop result~\cite{susydelrhos,susydelrhol}.
It is remarkable that contrary to the Standard Model case~\cite{lb},
$
\Delta \rho ^{\rm SM}_1 = - \Delta \rho_0^{\rm SM} \,
\frac{2}{3} \frac{\alpha_s}{\pi} (1+ \frac{\pi^2}{3} ) ,
$
where the QCD corrections are negative and screen the one-loop result,
$\Delta \rho ^{\rm SUSY}_{1, {\rm gluon}}$ enters with the same sign as
the one-loop contribution. It therefore enhances the sensitivity in the
search for the virtual effects of scalar quarks in high-precision
electroweak measurements.

The analytical expression for the gluino-exchange contribution is much
more complicated than 
for gluon-exchange.
In general the gluino-exchange
diagrams give smaller contributions compared to gluon exchange.
Only for gluino and squark masses close to the experimental
lower bounds they compete with the gluon-exchange contributions.
In this case, the gluon and gluino contributions add up to about 30\%
of the one-loop value for maximal mixing~\cite{susydelrhos}.
For higher values of $m_{\tilde{\mathrm g}}$, the contribution
decreases rapidly since the gluino decouples in the large-mass limit.

\section{Gluonic corrections to $\De r$}

The leading contribution to $\De r$ in the MSSM can be approximated by
the contribution to the $\rho$~parameter according to
$
\De r \approx - \cw^2/\sw^2 \De \rho,
$
where $\cw^2 = 1 - \sw^2 = \MW^2/\MZ^2$.
In order to test the accuracy of this approximation, we have derived the
exact result for the gluon-exchange 
correction to the contribution of a squark doublet to $\De r$.
It
is given by
\beq
\Delta r^{\rm SUSY}_{\rm gluon} = \Pi^{\gamma}(0) -
\frac{\cw^2}{\sw^2} \left(\frac{\de \MZ^2}{\MZ^2} -
\frac{\de \MW^2}{\MW^2} \right) +
\frac{\Sigma^\PW(0) - \de \MW^2}{\MW^2},
\label{eq:deltrglu}
\eeq
where $\de \MW^2 = \Re \Sigma^{\PW}(\MW^2)$,
$\de \MZ^2 = \Re \Sigma^{\PZ}(\MZ^2)$, and $\Pi^{\gamma}$,
$\Sigma^{\PW}$, and $\Sigma^{\PZ}$ denote the transverse parts of the
two-loop gluon-exchange contributions to the photon vacuum polarization
and the W-boson and Z-boson \ses, respectively,
which all are understood to contain the subloop renormalization.

\begin{figure}[htb]
\begin{center}
\mbox{
\psfig{figure=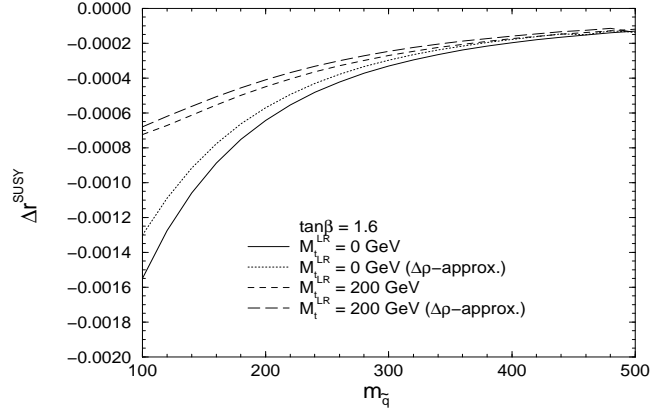,width=8.5cm,height=5.5cm,
              bbllx=25pt,bblly=85pt,bburx=550pt,bbury=445pt}}
\parbox{10cm}{
\caption[]{Contribution of the gluon-exchange diagrams to
$\Delta r^{\rm SUSY}$.
The exact result is compared with the $\De\rho$ approximation.
\label{fig:deltrglu}
}}
\end{center}
\end{figure}

The gluon-exchange correction to the contribution of the stop/sbottom
loops
to $\De r$ is shown in \reffi{fig:deltrglu} together with the
$\De\rho$ approximation,
$\De r \approx - \cw^2/\sw^2 \De \rho$,
as a function of the common
scalar mass parameter
$
m_{\tilde q} =
M_{\tilde{\Pt}_{L}}=M_{\tilde{\Pt}_{R}}=M_{\tilde{\Pb}_{L}} =
M_{\tilde{\Pb}_{R}} ,
$
where the $M_{\tilde{q}_i}$ are the soft SUSY breaking parameters
appearing in the stop and sbottom mass matrices
as specified in \citere{susydelrhol}.
In this scenario, the
scalar top mixing angle is either very small, $\theta_{\tilde \Pt} \sim
0$, or almost maximal, $ \theta_{\tilde \Pt} \sim  -\pi/4$, in most of
the MSSM parameter space. The plots are shown for the two cases
$M^{LR}_\Pt=0$ (no mixing) and $M_\Pt^{LR}=200$~GeV (maximal mixing) for
$\tb=1.6$.

The two-loop contribution $\Delta r^{\rm SUSY}_{\rm gluon}$ is of the
order of 10--15\% of the one-loop result. It yields a shift in
the W-boson mass of up to $20$~MeV for low values of $m_{\tilde q}$ in
the no-mixing case. If the parameter $M^{LR}_\Pt$
is made large or the 
assumption of a common scalar mass parameter
is relaxed, much
bigger effects are possible~\cite{susydelrhol}.
As can be seen in \reffi{fig:deltrglu}, the $\De\rho$ contribution
approximates the full result rather well. The two results agree within
10--15\%.


\section{Conclusions}

The two-loop ${\cal O}(\alpha_s)$ corrections to the $\rho$~parameter
has been derived in the MSSM. 
The gluonic corrections are of ${\cal O}(10\%)$: they are positive and
increase the sensitivity in the search for scalar quarks through their
virtual effects in high-precision electroweak observables. The gluino
contributions are in general smaller except for relatively light
gluinos and scalar quarks; the contribution vanishes for large gluino
masses. The exact result for the gluon-exchange correction to the
contribution of squark loops to $\De r$ has also been presented. It
gives rise to a shift in the W-boson mass of up to $20$~MeV. The result
has been compared with the leading contribution entering via the
$\rho$~parameter, and good agreement has been found.

\smallskip

The author thanks A.~Djouadi, P.~Gambino, S.~Heinemeyer, W.~Hollik and
C.~J\"unger for collaboration on this subject.

%


\begin{thebibliography}{99}

\frenchspacing
\newcommand{\anp}[3]{{\sl Ann.~Phys.} {\bf #1} (19#2) #3}
\newcommand{\app}[3]{{\sl Acta~Phys.~Pol.} {\bf #1} (19#2) #3}
\newcommand{\cmp}[3]{{\sl Commun. Math. Phys.} {\bf #1} (19#2) #3}
\newcommand{\cpc}[3]{{\sl Comp. Phys. Commun.} {\bf #1} (19#2) #3}
\newcommand{\fop}[3]{{\sl Fortschr. Phys.} {\bf #1} (19#2) #3}
\newcommand{\ijmp}[3]{{\sl Int. J. Mod. Phys.} {\bf #1} (19#2) #3}
\newcommand{\jetp}[3]{{\sl JETP} {\bf #1} (19#2) #3}
\newcommand{\jetpl}[3]{{\sl JETP Lett.} {\bf #1} (19#2) #3}
\newcommand{\jmp}[3]{{\sl J. Math. Phys.} {\bf #1} (19#2) #3}
\newcommand{\mpl}[3]{{\sl Mod. Phys. Lett.} {\bf #1} (19#2) #3}
\newcommand{\nc}[3]{{\sl Nuovo Cimento} {\bf #1} (19#2) #3}
\newcommand{\nim}[3]{{\sl Nucl. Instr. Meth.} {\bf #1} (19#2) #3}
\newcommand{\np}[3]{{\sl Nucl. Phys.} {\bf #1} (19#2)~#3}
\newcommand{\npB}[3]{{\sl Nucl. Phys.} {\bf B #1} (19#2)~#3}
\newcommand{\nphbps}[3]{{\sl Nucl. Phys.} {\bf B} {\it (Proc. Suppl.)}
{\bf #1B} (19#2) #3}
\newcommand{\plB}[3]{{\sl Phys. Lett.} {\bf B #1} (19#2) #3}
\newcommand{\prD}[3]{{\sl Phys. Rev.} {\bf D #1} (19#2) #3}
\newcommand{\prl}[3]{{\sl Phys. Rev. Lett.} {\bf #1} (19#2) #3}
\newcommand{\pl}[3]{{\sl Phys. Lett.} {\bf #1} (19#2) #3}
\newcommand{\ptp}[3]{{\sl Prog. Theo. Phys.} {\bf #1} (19#2) #3}
\newcommand{\sptp}[3]{{\sl Suppl. Prog. Theo. Phys.} {\bf #1} (19#2) #3}
\newcommand{\sjnp}[3]{{\sl Sov. J. Nucl. Phys.} {\bf #1} (19#2) #3}
\newcommand{\zp}[3]{{\sl Z. Phys.} {\bf #1} (19#2) #3}
\newcommand{\vj}[4]{{\sl #1~}{\bf #2} (19#3) #4}
\newcommand{\ej}[3]{{\bf #1} (19#2) #3}
\newcommand{\vjs}[2]{{\sl #1~}{\bf #2}}

\bibitem{susyprec}
P.~Chankowski, A.~Dabelstein, W.~Hollik, W.~M\"osle, S.~Pokorski
and J.~Rosiek, \np{B 417}{94}{101};\\
D.~Garcia and J.~Sol\`a, \mpl{A 9}{94}{211};\\
D.~Garcia, R.~Jim\'enez, J.~Sol\`a, {\sl Phys. Lett.} {\bf B 347}
(1995) 309 and 321;\\ 
D.~Garcia and J.~Sol\`a, {\sl Phys. Lett.} {\bf B 357} (1995) 349;\\
P.~Chankowski and S.~Pokorski, {\sl Nucl. Phys.} {\bf B 475} (1996) 3;\\
W.~de Boer, A.~Dabelstein, W.~Hollik, W.~M\"osle and U.~Schwickerath,
\zp{C 75}{97}{627}.

\bibitem{susydelrhos}
A.~Djouadi, P.~Gambino, S.~Heinemeyer, W.~Hollik, C.~J\"unger and
G.~Weiglein, \prl{78}{97}{3626}.

\bibitem{susydelrhol}
A.~Djouadi, P.~Gambino, S.~Heinemeyer, W.~Hollik, C.~J\"unger and
G.~Weiglein, KA-TP-8-1997, hep-ph/9710438.

\bibitem{R6rho}
R.~Barbieri and L.\ Maiani, \np{B 224}{83}{32}; \\
M.~Drees and K.~Hagiwara, \prD{42}{90}{1709}.

\bibitem{velt}
M.~Veltman, \np{B 123}{77}{89}.

\bibitem{lb}
A.~Djouadi and C.~Verzegnassi, \pl{B 195}{87}{265};\\
A.~Djouadi, \nc{A 100}{88}{357}.

\end{thebibliography}
\end{document}